\documentclass[fleqn,usenatbib]{mnras}
\usepackage{newtxtext,newtxmath}
\usepackage[T1]{fontenc}

\DeclareRobustCommand{\VAN}[3]{#2}
\let\VANthebibliography\thebibliography
\def\thebibliography{\DeclareRobustCommand{\VAN}[3]{##3}\VANthebibliography}

\usepackage{graphicx}	% Including figure files
\usepackage{amsmath}	% Advanced maths commands
\usepackage{hyperref}

\usepackage{subcaption}
\title[VY Scl Accretion States]{Characterising high and low accretion states in VY Scl CVs using ZTF and TESS data }
\author[C. Duffy et al.]{
C.~Duffy,$^{1,2,3}$\thanks{Contact e-mail: \href{c.j.duffy@lancaster.ac.uk}{c.j.duffy@lancaster.ac.uk}}
Kinwah Wu,$^{4}$
G.~Ramsay,$^{2}$
Matt A. Wood,$^{5}$
Paul A. Mason,$^{6,7}$
Pasi Hakala,$^{8}$
D. Steeghs$^{3}$
\\
$^{1}$Department of Physics, Lancaster University, Lancaster, LA1 4YB, UK\\
$^{2}$Armagh Observatory and Planetarium, College Hill, Armagh, BT61 9DB, Northern Ireland, UK\\
$^{3}$Department of Physics, University of Warwick, Gibbet Hill Road, Coventry, CV4 7AL, UK\\
$^{4}$Mullard Space Science Laboratory, University College London, Holmbury St. Mary, Surrey RH5 6NT, UK\\
$^{5}$Department of Physics \& Astronomy, Texas A\&M University--Commerce, Commerce, TX, 75429-3011, USA\\
$^{6}$New Mexico State University, MSC 3DA, Las Cruces, NM, 88003, USA\\
$^{7}$Picture Rocks Observatory, 1025 S. Solano Dr. Suite D., Las Cruces, NM 88001, USA\\
$^{8}$Finnish Centre for Astronomy with ESO (FINCA), Quantum, Vesilinnantie 5, FI-20014, University of Turku, Finland\\
}

\date{Accepted 2024 October 24. Received 2024 October 18; in original form 2024 July 5}

\pubyear{2024}

\begin{document}
\label{firstpage}
\pagerange{\pageref{firstpage}--\pageref{lastpage}}
\maketitle

\begin{abstract}
VY Scl binaries are a sub-class of cataclysmic variable (CV) which show extended low states, but do not show outbursts which are seen in other classes of CV. To better determine how often these systems spend in low states and to resolve the state transitions we have analysed ZTF data on eight systems and {\sl TESS} data on six systems. Half of the sample spent most of the time in a high state; three show a broad range and one spends roughly half the time transitioning between high and low states. Using the ZTF data we explore the colour variation as a function of brightness. In KR Aur, we identify a series of repeating outburst events whose brightness appears to increase over time. Using {\sl TESS} data we searched for periods other than the orbital. In LN UMa we find evidence for a peak whose period varies between 3--6 d. We outline the current models which aim to explain the observed properties of VY Scl systems which includes disc irradiation and a white dwarf having a significant magnetic field.

\end{abstract}


%%%%%%%%%%%%%%%%% BODY OF PAPER %%%%%%%%%%%%%%%%%%
\section{Introduction}

Cataclysmic variables (CVs) are interacting binaries with a white dwarf (WD) accreting material from, usually, a late type (donor) star \citep[see][for a review]{warner2003cataclysmic}. Mass transfer in CVs occurs via Roche-lobe overflow from the donor star through the inner Lagrangian (L1) point of the binary. If the WD is modestly magnetised $(B \lesssim0.1$ MG), an accretion disc can be formed, mediating material inflow onto the WD via a boundary layer. CVs often show high and low optical states. Except for Polars, which are discless CVs containing a highly magnetic WD, the two states correspond respectively to a hot, bright disc with high viscosity and a cool, faint disc with a low viscosity \citep[see][]{1993adcs.book....6C}. According to the Disc Instability Model \citep[DIM; for a review see][]{2020AdSpR..66.1004H}, the existence of state bimodality and state transitions are associated with CV outbursts, where one or more physical parameters of the accretion disc changes in such a manner as to make the disc unstable, which results in an abrupt increase in the mass accretion onto the WD. The long-term behaviour of mass transfer in CVs and the (in)stability of their accretion discs provides insight into the accretion and mass transfer processes in general and the nature of viscosity that drives the accretion flow \citep{warner2003cataclysmic}.

VY Scl binaries, also known as ``anti-dwarf novae'', are a sub-class of the non-magnetic CVs. They are distinguished from nova-like systems, which have a high mass transfer rate, as their optical brightness can decrease by several magnitudes for up to several hundred days at a time \citep{warner2003cataclysmic}. The drop in the brightness reflects the decrease in the accretion flow onto the WD, which is attributed to the decrease in the mass-transfer rate, with possibly even a temporary cessation of mass transfer from the secondary star in some systems. Noticeably, VY Scl binaries have not been observed in outburst, despite the DIM predicting that outbursts should occur in their low state \citep{1999MNRAS.305..225L}. However, lower amplitude ``stunted" outbursts have been observed in some VY Scl binaries during their high states \citep[e.g.,][]{2014AJ....147...10H}. 

Various models have been proposed to explain the occurrence of low states in CVs, with star-spot models amongst them. In semi-detached binaries such as CVs, where the rate of mass transfer is determined by the physical condition of the low-mass donor star at the L1 point, the gas temperature and the atmospheric pressure scale heights are two core parameters \citep{Lubow1975ApJ}. Low-mass stars are often magnetically active, with some showing coronal flares and also migrating star-spots. The passing of star-spots to regions close to the L1 point would lead to a reduction in gas temperatures and atmospheric pressure scale height, which in turn modifies the mass transfer rate \citep[see][]{1994ApJ...427..956L}. 
  
The star-spot models highlight the importance of stellar magnetism in regulating the mass transfer in low-mass semi-detached binaries, and it is an starting point for building more holistic models, such as those invoking a magnetic valve mechanism  \citep[e.g.][]{Wu2008,2022MNRAS.516.3144D,2024ApJ...965...96M}, where the magnetic interactions between the WD and the companion star control mass transfer in the binary. This provides a unified description for high and low states, the state transition and the duty cycles in some CV subclasses, in particular the Polars, covering both systems with low and high WD magnetic fields
 \citep[$B\sim$10-200 MG;][]{2024ApJ...965...96M}. 
 
While there is evidence that the magnetic valve mechanism can explain high/low states in Polars \citep{2022ApJ...938..142M}, in the same form it is clearly inapplicable to the non-magnetic VY Scl binaries. However, such a scenario in VY Scl binaries would rely solely on the atmospheric magnetic activity of the low-mass companion star. We note that what drives the state transition, sustains the mass transfer during the high state and halts the mass transfer during the low state would also depend additionally on the mechanical and thermal conditions of the low-mass donor star, as well as the orbital dynamics of the binary.  

With the exception of Polars, CVs possess an accretion disc. During the low state the accretion disc is expected to be cool. It may evolve in such a way that a disc instability develops, triggering outbursts \citep[see e.g.][]{1993adcs.book....6C}. The lack of outbursts observed in VY Scl binaries suggests that some mechanism(s) are present which inhibit disc instability. \citet{1999MNRAS.305..225L} proposed that the anticipated outbursts are suppressed by a high WD atmospheric temperature ($\sim$40,000 K), which heats the inner disc and thus prevents the development of the instability needed for outburst. This proposition is consistent with observations of the WDs in some VY Scl binaries which are found to be relatively hot \citep[see e.g.][]{Mizusawa2010PASP}. Whilst it has explained the absence of outbursts in the low state, it is yet to account for the lack of outbursts in the transition between the low and high states, known as the intermediate state,  \citep[see e.g.][]{1999MNRAS.305..225L,Zemko2014MNRAS}, which is observed in some systems. The durations of these intermediate states are longer than the viscous timescale, and as the accretion disc is in an unstable configuration, the DIM would predict outbursts to occur.

In order to explain the general absence of outbursts, \citet{2002A&A...394..231H,2005astro.ph..6382H} proposed that the magnetic field of the WDs in VY Scl binaries are strong enough to affect the accretion flow. They showed that a relatively weak magnetic field ($B \sim40$~kG) is sufficient to prohibit the formation of a low-state accretion disc around a WD of $\sim0.7\ M_{\sun}$. For a similar system, a WD magnetic field $B \gtrsim 6$~MG would be sufficient to disrupt the disc during the majority of the intermediate state so that outbursts would not be possible. Such a WD magnetic field strength would place the binaries into the CV subclass Intermediate Polars, whose WDs generally have a magnetic field  of  $10\;\! {\rm MG} \gtrsim  B \gtrsim 0.1\;\! {\rm MG}$.

The absence of a full-scale accretion disc, proposed by \citet{2002A&A...394..231H,2005astro.ph..6382H}, was used to explain the absence of outbursts in the low state of VY Scl binaries. A more extreme proposal is that the accretion disc is entirely absent in the low state due to a complete cessation of mass transfer. Whilst the enhanced magnetic fields associated with star spots migrating to the L1 region can temporarily halt the mass outflow from the donor star, it is hard to sustain for a prolonged time unless an external force e.g. the WD magnetic field in the magnetic valve scenario for Polars described in \cite{2022MNRAS.516.3144D} and \citet{2024ApJ...965...96M} is present to prevent the star spots migrating out of the L1 region. However, if the mass transfer process is intrinsically unstable, the binary can enter a prolonged low state where mass transfer ceases and a full-scale accretion disc is naturally absent. The study of \citet[][]{Wu1995PASA} has shown that irradiative heating feedback can trigger such mass-transfer instabilities. 
  
Attempts have been made to verify the star-spot and the magnetic disc disruption scenarios, but the situation is complicated by the fact that whilst some observations support one or other of the scenarios, other observations disprove them. The study of MV Lyr by \citet{Linnell2005ApJ}, for example, suggests that neither the star-spot nor the disc-disruption scenario is is applicable with the WD and/or the inner disc being too cool and also showing no evidence of a WD magnetic field of sufficient strength. On the other hand, the absence of the accretion disc in TT Ari \citep{1999A&A...347..178G} would favour the magnetic disc-disruption scenario. {Similarly, \citet{2023MNRAS.521.5846M} found that during low states SDSS J154453.60+255348.8 did not show evidence for an accretion disc, which supports the predictions of \citeauthor{2002A&A...394..231H}.} However, when looking at VY Scl, the sub-class archetype, \citet{2018A&A...617A..16S} found evidence of an accretion disc during the low state, and they also used this to rule out the disc-disruption scenario.

Despite VY Scl binaries generally not outbursting during their low states, ``stunted" outbursts were identified by \citet{2004AJ....128.1279H} in FY Per and V794 Aql during their low states. These stunted outbursts have greatly reduced amplitude  compared with those which would be expected from an outburst in similar non-VY Scl systems. \citet{2018A&A...617A..16S} also found aperiodic variations in the brightness of VY Scl itself, and suggested that they are a similar phenomenon. The nature of stunted outbursts is unclear: they could be associated with certain disc instabilities, or alternatively fluctuations in mass transfer rates, or a combination of both \citep{1998AJ....115.2527H,2001PASP..113..473H}. These outbursts require the accretion disc to act as an agent, and thus they would exclude the possibility of a strongly magnetised WD. Magnetically gated outbursts have also been observed in MV Lyr \citep{2017Natur.552..210S} from which they inferred a WD magnetic field of similar magnetic to those proposed by \citet{2002A&A...394..231H,2005astro.ph..6382H}.

In this paper we investigate the long-term behaviour of eight VY Scl binaries using the multi-colour photometric data from the Zwicky Transient Facility \citep[\textit{ZTF};][]{2019PASP..131a8002B} in order to better understand the long term behaviour of eight VY Scl binaries and therefore the underlying processes that give rise to the observed high and low accretion states, and their duty cycles. We use also extended optical photometric data from the Transiting Exoplanet Survey Satellite  \citep[\textit{TESS};][]{2015JATIS...1a4003R} to search for periods longer than the orbital period and resolve transitions between high and low states.

%We use data from the Zwicky Transient Facility (\textit{ZTF}) and the Transiting Exoplanet Survey Satellite (\textit{TESS}) to probe the behaviour of a number of known VY Scl systems in order to improve our understanding of the processes underpinning this subclass of CVs.

\begin{table*}
    \centering
    \caption[Properties of considered VY Scl systems]{List of sources analysed in this work, detailing the feature of interest to this work and the origin(s) of the data considered for the analysis. Features; transitions, ``outburst'', and superorbital periods are denoted by T, O and S respectively.}
    \label{tab:overview}
    \begin{tabular}{lcccccc}
    \hline
    Source & Orbital Period [min] &Feature(s) & \% High State & Mean Low State Depth [mag] &\textit{ZTF} & \textit{TESS}\\
    \hline
    RX J2338+431  & $187.7^{a}$ & T    & $93.5\%$           &3.1&\checkmark  &\checkmark \\
    MV Lyr        & $191.4^{b}$ & T    & $81.0\%^{\dagger}$ &5.0&\checkmark  &\checkmark \\
    TT Ari        & $198.1^{c}$ & S    & N/A                &N/A&$\times$    &\checkmark \\
    LN UMa        & $207.9^{d}$ & O, P & N/A                &N/A&\checkmark  &\checkmark \\
    V794 Aql      & $220.8^{e}$ & T, O & $83.1\%^{*}$       &2.8&\checkmark  &$\times$   \\
    BZ Cam        & $221.3^{f}$ & T, O & $90.2\%$           &0.7&\checkmark  &\checkmark \\
    ES Dra        & $225.6^{g}$ & T, O & $90.0\%$           &1.4&\checkmark  &$\times$ \\
    KR Aur        & $234.4^{h}$ & T, O & $35.4\%$           &4.6&\checkmark  &$\times$   \\
    V504 Cen      & $252.8^{i}$ & S    & N/A                &N/A&$\times$    &\checkmark \\
    MP Gem        & Unknown     & T    & $74.4\%^{\dagger}$ &4.0&\checkmark  &$\times$   \\
    \hline
    \end{tabular}\\
\begin{flushleft}    
\textit{Notes. Citations:} \footnotesize{$^{a}$\citet{2018AJ....156..231W}, $^{b}$\citet{1995PASP..107..545S}, $^{c}$\citet{2002ApJ...569..418W}, $^{d}$\citet{1998AJ....115.2044H}, $^{e}$ \citet{1998AJ....115.2527H}, $^{f}$\citet{1996AJ....111.2422P}, $^{g}$\citet{2012NewA...17..108R}, $^{h}$\citet{1983ApJ...267..222S}, $^{i}$\citet{2022MNRAS.514.4718B} }\\
\footnotesize{$^{*}$} The high state in V794 Aql was defined as those points brighter 1 mag below the mean high state brightness, in order to account for the high brightness variability seen in the high state of this system. Using the definition applied to other systems yields a value of 49.3\%\\
\footnotesize{$^{\dagger}$} These systems have visibility gaps during the majority of the suspected low state times, leading to artificially inflating the \% high state. {Estimated points filling in the established low state at the same cadence as the observations were inserted for these systems in order to calculate these values, without these estimated points} the values are MP Gem: 91.3\% and MV Lyr: 88.7\%.\\
N/A denotes systems where no low states are seen in the data considered, however, as known VY Scl systems it would be incorrect to record a 100\% high state.
\end{flushleft}
\end{table*}

\begin{figure*}
    \includegraphics[width=\textwidth]{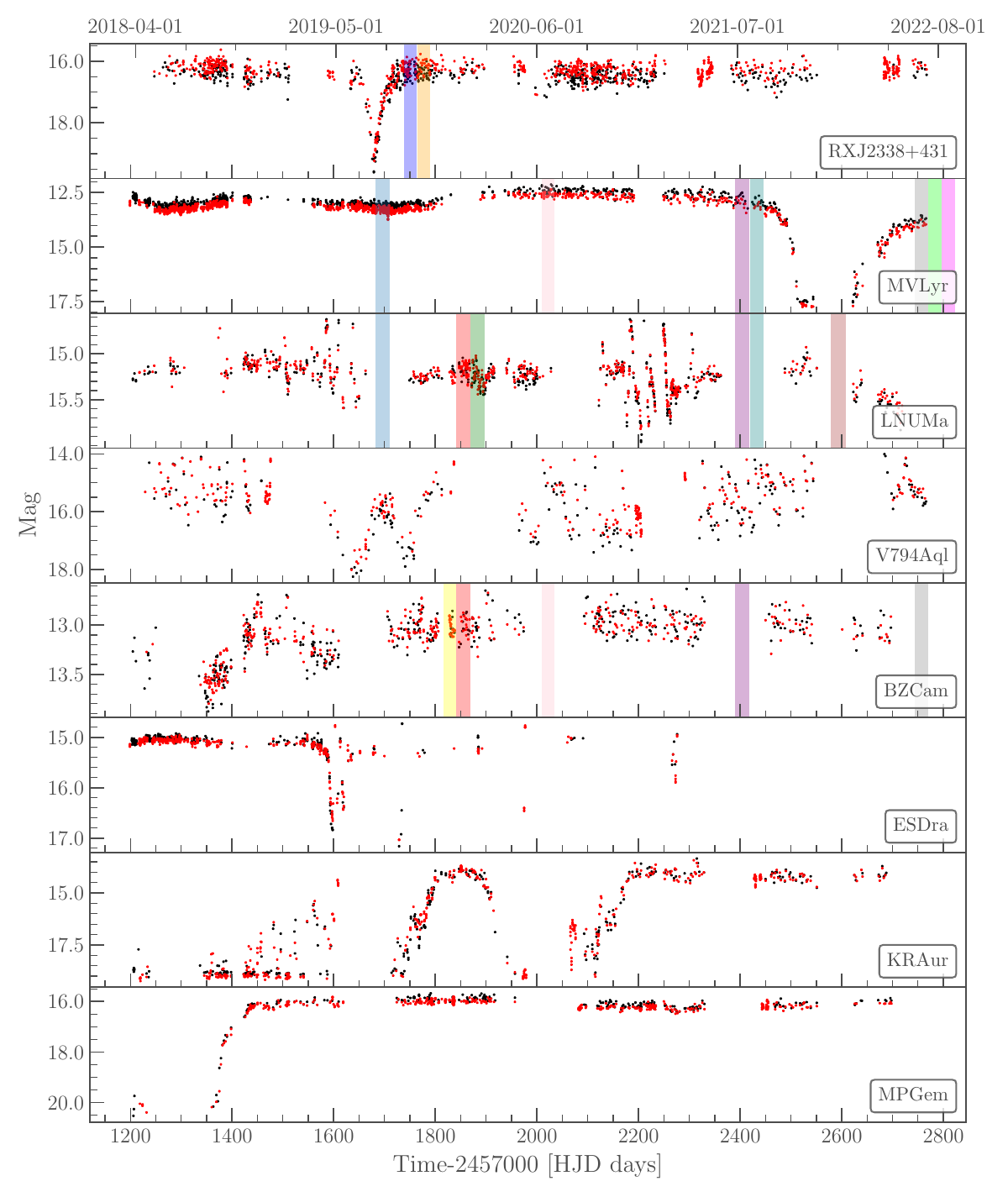}
    \caption{The \textit{ZTF} lightcurves of the 8 sources which were found to show features of interest. Black and red data points denote ZTF-\textit{g} and ZTF-\textit{r} data respectively. The windows denote the times of \textit{TESS} observations with colours indicating the sector; blue - s14, dark blue - s16, orange - s17, yellow - s19, red - s20, green - s21, pink - s26, purple - s40, teal - s41, brown - s-47, gray - s53, lime - s54, fuchsia - s55. Group A systems: RX~J2338+432, MV~Lyr, ES~Dra, KR~Aur and MP~Gem. Group B systems: LN~UMa, V794~Aqr and BZ~Cam. (see \textsection\ref{sec:di-morphology}).}
\label{fig:example}
\end{figure*}

\section{Photometric Data}

We searched the \textit{ZTF} and \textit{TESS}\footnote{The sources in this paper were included on the 20 sec or 2 min cadence list thanks to their inclusion on the following Guest Investigator programmes: G022071/PI Scaringi; G022208/PI Zakamska; G022116/PI Schlegel; G022141/PI Barlow; G022254/PI Sion; G03071/PI Scaringi; G03044/PI Scaringi; G04208/PI Littlefield; G04046/PI Scaringi.}  lightcurves of all known VY Scl systems and found 10 systems which showed features such as transitions or short duration brightenings, as detailed in \autoref{tab:overview}\footnote{We initially accessed photometry on 23 known or suspected VY Scl systems, those which did not show events of interest pertinent to this work were excluded from further consideration This included BH Lyn and LX Ser in which eclipses were visible in the \textit{TESS} data. }. The \textit{ZTF} data on these systems, shown in \autoref{fig:example}, was collected to allow us to probe the longer term behaviour seen in these systems, quantifying the occurrence and duration of the features of interest. The \textit{TESS} photometry of these systems allows us to probe their short term behaviour, which gives us a more precise understanding of the processes at work in these features. 

\textit{ZTF} is a time domain survey program using the Palomar 1.2 m telescope, which is capable of covering 47\degr\textsuperscript{2} per exposure. As such \textit{ZTF} is capable of surveying the entire sky rapidly, providing photometry on a diverse number of sources with a limiting brightness of $\rm mag \sim20$. \textit{ZTF} operates a number of different survey programs including the public survey which accounts for 50\% of the observation time. The public survey makes observations in two filters ZTF-\textit{g} and ZTF-\textit{r}, $\sim$4087--5522 \AA\ and $\sim$5600--7316 \AA\ respectively \citep[see][for a full description of \textit{ZTF} operations]{2019PASP..131a8002B}. {We downloaded ZTF photometry from the NASA/IPAC archive\footnote{\href{https://irsa.ipac.caltech.edu/Missions/ztf.html}{https://irsa.ipac.caltech.edu/Missions/ztf.html}}. The data used has been processed through the standard ZTF pipeline, where each photometric point is assigned a quality flag, \texttt{catflags}. Only data with \texttt{catflags=0} were included in analysis.}

\textit{TESS} is a space observatory launched in 2018. It makes observations of pre-defined sectors using four wide field telescopes, each with a $24\degr \times 24\degr$ field of view . 
%Each of the 26 sectors span from the ecliptic plane to the ecliptic pole. 
Each sector lasts for 27 days and is near continuous, with a short mid-point break for data transmission and momentum unloading, and generates 20-s (from the first extended mission onwards) and 120-s cadence photometry on a list of predefined targets \citep[see][for full details of the \textit{TESS} mission]{2015JATIS...1a4003R}. {We downloaded the calibrated lightcurves of all sources from the MAST data archive\footnote{\href{https://archive.stsci.edu/tess/}{https://archive.stsci.edu/tess/}}. We used the data values for {\tt PDCSAP\_FLUX}, which are the Simple Aperture Photometry values, {\tt SAP\_FLUX}, after the removal of systematic trends common to all stars in that chip. Each photometric point is assigned a \texttt{QUALITY} flag which indicates if the data has been compromised to some degree by instrumental effects. We removed those points which did not have \texttt{QUALITY=0} and normalised each light curve by dividing the flux of each point by the mean flux of the  star.}

\subsection{Analysis}

\begin{figure*}
    \includegraphics[width=\textwidth]{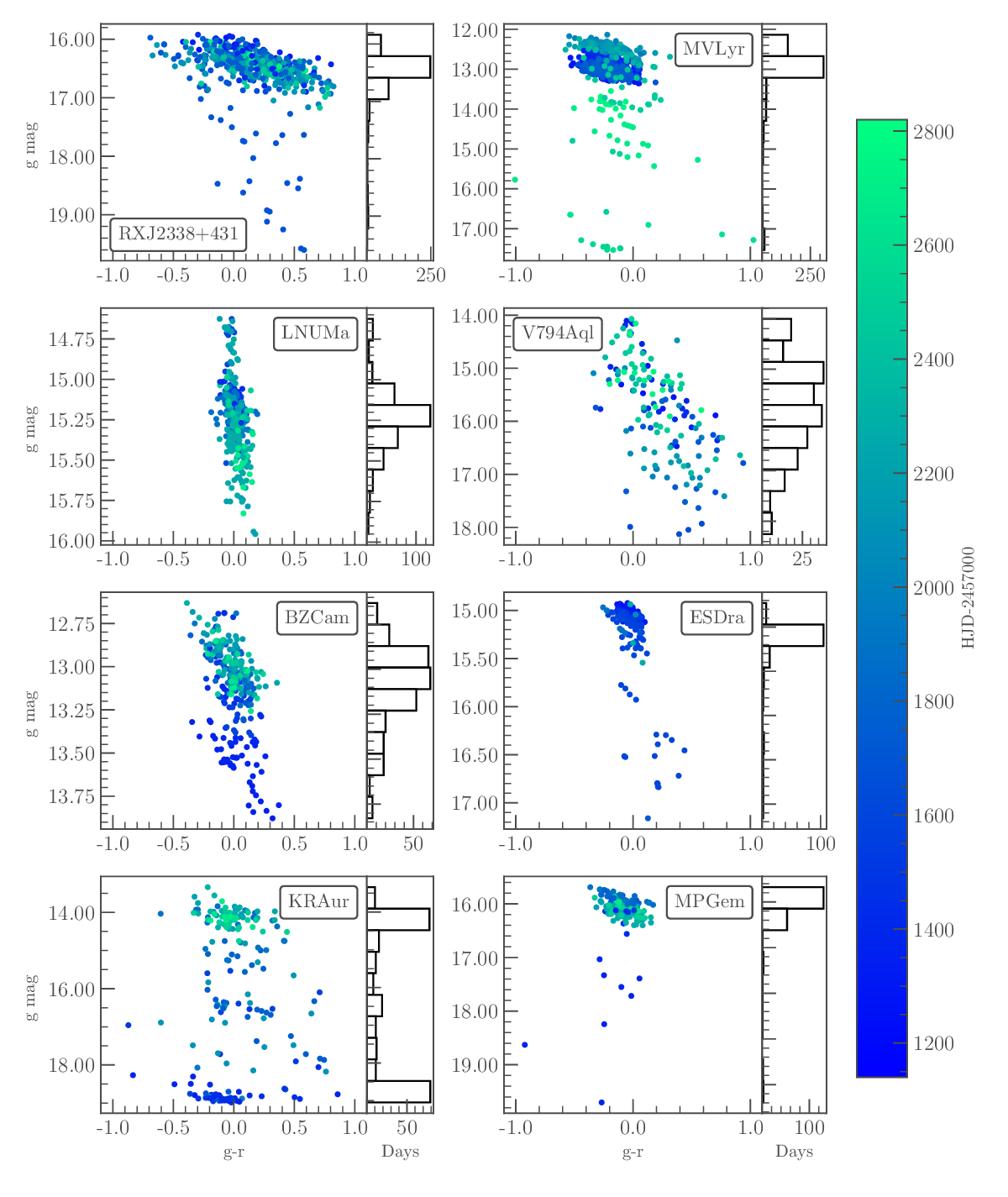}
    \caption{The \textit{ZTF} colour-magnitude diagrams of each of the \textit{ZTF} sources. Each datum has been coloured according to the associated \textit{g}-band observation time using the same time range as seen in \autoref{fig:example}. Attached to each colour magnitude is a histogram showing the number of days in which observations in a given \textit{g}-band brightness bin were made. {Errors on the magnitude and colour index were calculated, though in almost every case were contained entirely within the size of the marker.}}
    \label{fig:colourDisto}
\end{figure*}

To investigate the variation in state behaviour and transitions {over the duration of observations} we quantified the distribution of the brightness states in the systems where we saw both high and low states. Histograms showing the number of days a given brightness bin occurs in the \textit{ZTF g}-band lightcurves are shown as part of \autoref{fig:colourDisto}. Although this allows for a good visual interpretation of the distribution of brightness states it does have limitations. It is heavily dependent on the sampling of the different states, e.g., for MV Lyr, although the lightcurve indicates the presence of a $\sim$200~d low state as this occurs during a visibility gap it is underrepresented by the histogram. Despite this, we used these to quantify the fraction of time that a system occupies a high state as shown in \autoref{tab:overview}. This was completed by visually identifying a section of a system's lightcurve which corresponded the high state and using this section to determine a mean high state brightness. The high state was then defined to be when the {lightcurve was no fainter than 0.5 mag below the} mean high state brightness. Using 1D histograms (although with 100 bins instead of 10 bins as shown in \autoref{fig:colourDisto}) we were subsequently able to determine the {distribution of brightness within the lightcurve and thus the} fraction of time each binary spent in a high state. This shows that for most cases in which we see a transition it spends $\sim$90\% of the time in a high state, except KR Aur where the duty cycle is only $\sim35$\%.

We have quantified the behaviour of the transitions seen in each of the \textit{ZTF} data sets by calculating the $e$-folding time value\footnote{$\tau$ is the value describing time required for the brightness to change by a factor of $e$ and is given by the expression ($\log_{10}e/0.4)/(\Delta m/\Delta t) = 1.086/(\Delta m/\Delta t)$ \citep{2004AJ....128.1279H}}, $\tau$, for each, which gives us a metric with which to describe the rate of change of brightness (or ``speed'') of the transition. We have tabulated the results of this calculation in \autoref{tab:eFolding}. Together with quantifying the duty cycle discussed above these metrics allow us to quantitatively discuss the nature of the different states and transitions seen in the VY Scl binary population.

\begin{table}
    \centering
    \caption{The $e$-folding time of each of the transitions in each of the sources seen to transition in \textit{ZTF}. Type refers to the direction of transition, with + denoting a transition from low to high state and - denoting a transition from high to low state. Times are given in HJD-2457000.}
    \label{tab:eFolding}
    \begin{tabular}{lcccc}
    \hline
    Source & Start [HJD] & End [HJD] &Type [+/-] &$\tau$ [d]\\
    \hline
    RX J2338+431    & 1650 & 1678 &-&12.2\\
                    & 1679 & 1720 &+&17.0\\
    MV Lyr          & 2401 & 2517 &-&50.8\\
                    & 2622 & 2765 &+&46.0\\
    V794 Aql        & 1582 & 1637 &-&24.9\\
                    & 1640 & 1693 &+&23.3\\
                    & 1694 & 1743 &-&32.1\\
                    & 1747 & 1792 &+&19.3\\
    BZ Cam          & 1343 & 1434 &+&160.8\\
    ES Dra          & 1573 & 1598 &-&17.0\\
    KR Aur          & 1716 & 1837 &+&24.6\\
                    & 1880 & 1956 &-&16.7\\
                    & 2112 & 2204 &+&25.0\\
    MP Gem          & 1125 & 1222 &-&33.5\\
                    & 1360 & 1444 &+&25.6\\
    \hline
    \end{tabular}\\
\begin{flushleft}    
\textit{Notes.} The first transition in MP Gem is measured with additional data from AAVSO.
\end{flushleft}
\end{table}

As \textit{ZTF} provides colour information from multi-band observations, it is possible to use these data in order to probe the colour evolution of these systems during different states or at singular moments in time. We show in \autoref{fig:colourDisto} colour-magnitude diagrams for each of the systems for which we have \textit{ZTF} data. In each the data have been coloured such that the time of observation can be compared with a source's parent lightcurve. This shows that many are well constrained in colour, with both low and high states recording similar colour indices with no apparent preferences to the time of observation. There are, however, some exceptions to this---e.g., V794 Aql shows a trend to increased redness as it becomes fainter.

In order to probe periodic features in both \textit{ZTF} and \textit{TESS} data we used Lomb-Scargle periodograms as developed by \texttt{Astropy} and implemented in \texttt{Lightkurve} \citep{1976Ap&SS..39..447L, 1982ApJ...263..835S, astropy:2013, astropy:2018, astropy:2022, 2018ascl.soft12013L}. Comparing periodograms allowed us to identify a range of periodic features present in both data sets, which gives us insight into the processes taking place. Although much of the \textit{TESS} data reveals periodic behaviour, it is often the known orbital period.
%, and as such we choose not to remark on it in this work. 

As can be seen in \autoref{fig:example}, there is evidence of a number of short duration brightness events taking place across various objects. As we see substantial variation in the manifestation of these events, our approach to studying them must reflect this individuality. Specifically, we consider the duration and recurrence time of these events, their temporal profile, and if possible their the colour index. This allows us to characterise these events and establish the physical processes which underpin them.

\subsection{Individual Systems}

Although our aim is to explore the features of VY Scl systems as a whole, we firstly discuss briefly the behaviour of individual systems. Here we focus on those systems observed by ZTF in \autoref{fig:example}, as these data illustrates the long term behaviour of the systems to be considered and where appropriate is compared and contrasted with shorter term, higher cadence observations from TESS. 

\subsubsection{RX J2338+431}
Initially seen in the previously reported high state \citep{2018AJ....156..231W}, the \textit{ZTF} data shows a single, short, low state, that does not plateau at a low state brightness, but instead almost immediately begins a transition back to the high state. The system remains at a brightness minimum for at most 5~d with the entire episode lasting at most 80~d. Although visually appearing similar to other events presented, the transitions in J2338 are some of the fastest we have observed. The high states outwith the episode are unremarkable with only a small ($\rm mag\sim0.5$) variation in brightness consistent with a combination of observational effects and orbital variability. J2338 does not show a robust variation in colour behaviour between low and high state---although there is a slight lean towards reddening in the low state, but the short nature of the transitions and low state make it difficult to confirm. Despite this, J2338 shows the greatest variation in colour index of any of the systems which we consider here.

The \textit{TESS} data available for this binary catches only the very end of the transition out of the low state and mostly covers the second high state. As such it is not a particularly useful resource for considering the low state or the transition therein. The coverage from \textit{TESS} indicates that the newly established high state is uneventful.

\subsubsection{MV Lyr}
In MV Lyr we observe a single low state at a brightness of $\rm mag\sim17.5$ which lasts $\sim$200~d: this is consistent with low states previously seen \citep{2017Natur.552..210S}, who also identified magnetically gated outbursts during low states in MV Lyr. The vast majority of the low state occurs during a visibility gap so we are unable to report on its features, including any possible missed outbursts. The transitions which we measure in MV Lyr are at the slower end of the measured distribution, in part this may be associated with the depth of the low state relative to the other system considered here. Outwith the low state we see MV Lyr in a high state where the brightness appears to vary over $\sim$400~d between $\rm mag\sim12.5$--13.5. Despite this, the shorter term variation on the order of a few days is minimal and entirely in line with expectations of a locally stable high state. MV Lyr shows a relatively narrow spread of colour indices which remains mostly consistent between the high and low states, perhaps becoming slightly redder through the low state. We additionally see some outliers as seen in \autoref{fig:colourDisto} but we believe that these can be discounted as spurious.

The \textit{TESS} observations of MV Lyr are almost entirely from the high state and provides a steady state lightcurve. Despite this \textit{TESS} does appear to confirm the long term variability seen by \textit{ZTF} with successive sectors varying in flux at a pattern that matches the \textit{ZTF} observations. 

\subsubsection{TT Ari}\label{sec:TTAri}

\begin{figure*}
	\includegraphics[width=\textwidth]{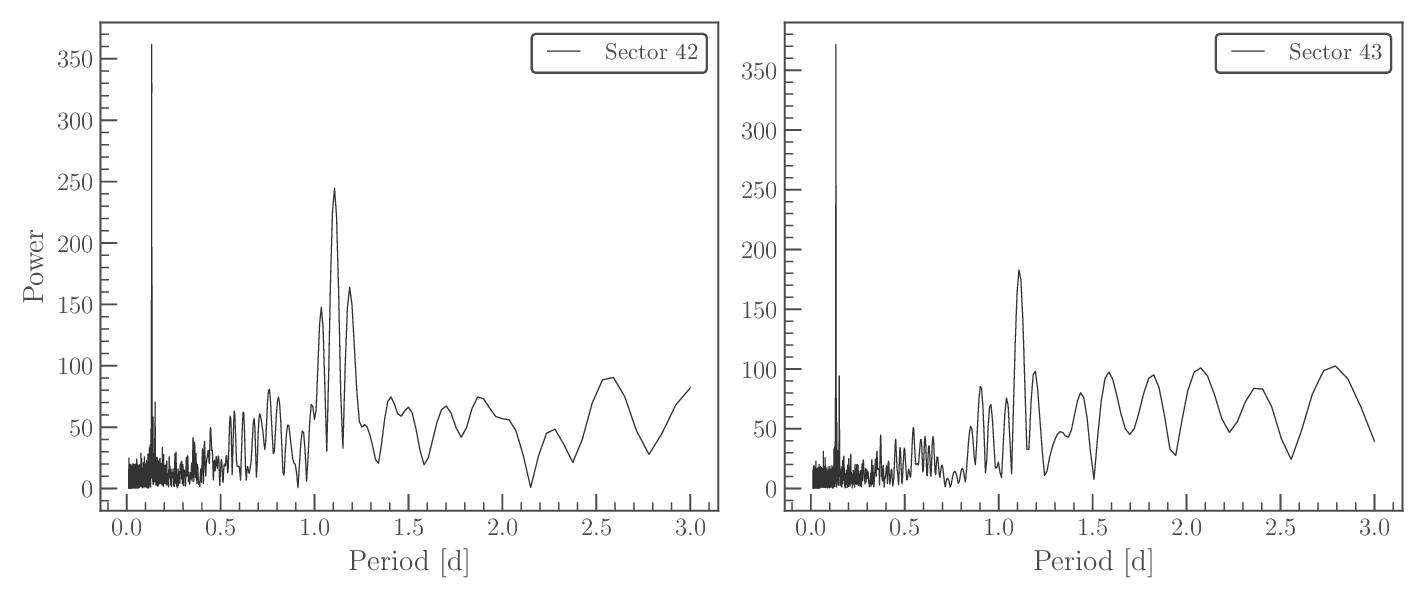}
	\caption[TESS periodograms of TT Ari]{The Lomb-Scargle periodograms of both of the TESS sectors of TT Ari observations showing the the beat (disc precession) period at 1.1~d, The strongest peak in both is the orbital period.}
	\label{fig:TTAriTESSPer}
\end{figure*}

{\textit{TESS} 20-s photometry of TT Ari is available in two successive sectors during the high state. In the power spectrum (\autoref{fig:TTAriTESSPer}) we observe the known orbital period variation (0.14~d), and periods of 0.15~d and 0.13~d which are the known positive and negative superhumps \citet{2022MNRAS.514.4718B}. We also see a longer term variation of $\sim$0.3 mag in the lightcurves. This was revealed in our period searching to be a 1.1~d period with a power between 0.5 and 0.75 $\rm d^{-1}$ that of the orbital period. This period was also also found by \citet{2022MNRAS.514.4718B} who identified it as a beat between the superhump periods which originates from the modulation of the negative superhump on the disc precession period. This precession period does not appear in the power spectra of this system but using Eqn.\ 1 in \citet{2009MNRAS.398.2110W} which relates the precession period to the positive superhump period and the orbital period we find an expected prograde precession period of 1.54~d. Long term study of TT Ari by \citet{2019MNRAS.489.2961B} identified a number of (quasi) periodic features, including a quasi-periodic oscillation during the high state, between 18 and 25 min; despite this we have been unable identify a similar feature in the \textit{TESS} data.}

\subsubsection{LN UMa}\label{sec:lnumaZTF}
Throughout the \textit{ZTF} observations, LN UMa is found in a high state with the majority of observations taking place at $\rm mag \sim 15.25$, which is the known high state brightness \citep{2009JAD....15....1P}. Despite this long running high state, we see a number of short-lived events wherein the brightness abruptly falls by as much as 0.75 mag in fewer than ten days. It then subsequently rises over $\sim$5~d to $\sim$0.5 mag brighter than the established highs state before dimming over $\sim$20~d (see \autoref{fig:example}). These events often occur in clusters with 4 or 5 seen in a $\sim$150~d time span, and as such one event may flow directly into the next. We believe that these events are the same erratic ``outbursting'' features that were identified by \citet{2004AJ....128.1279H}. We do not however, believe that these are true outbursts primarily because of the preceding drop in brightness and secondarily it would be unusual for a dwarf nova outburst of any sort to originate in the high state. LN UMa is static in colour throughout observations (see \autoref{fig:colourDisto}) with no significant change in colour index observed (mean $(g-r)=0.02$).

\begin{figure}
    \includegraphics[width=\columnwidth]{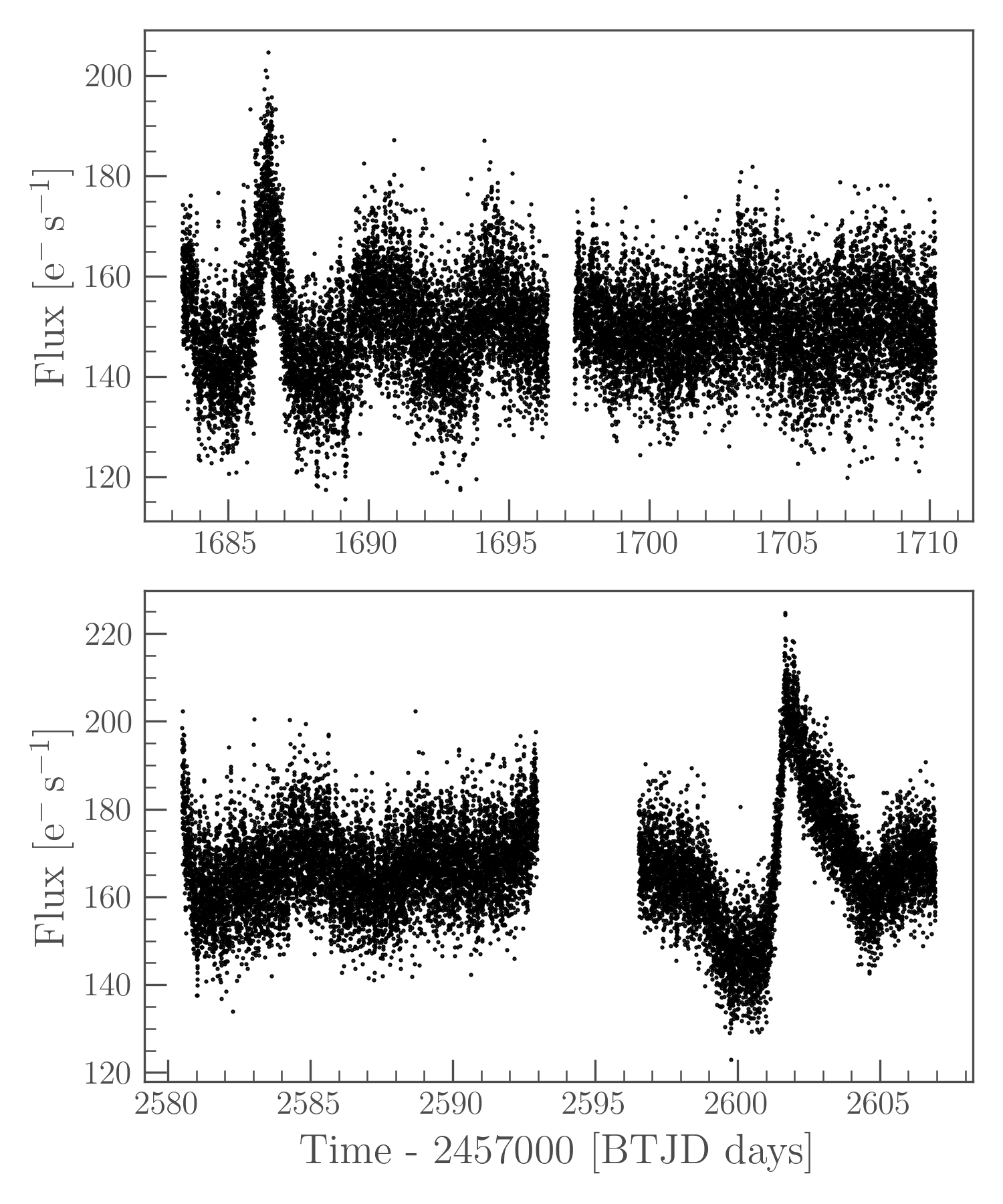}
    \caption{Extract of the \textit{TESS} lightcurve showing, upper: the sector 14 event and, lower: the sector 47 events in LN UMa.}
    \label{fig:LNUmaTESSevents}
\end{figure}

LN UMa was observed in six \textit{TESS} sectors with 120-s cadence contemporaneous with the \textit{ZTF} observations. In sectors 14 and 47 we see a $\sim$3~d increase in brightness; which in the sector 47 event is preceded by a decrease in brightness. The event in sector 14 has mostly symmetric temporal profile with an approximately equal rate of change of brightness in both the rise and the fall phases. The sector 47 event, in contrast, shows a rapid rise phase followed by a slower decline and appears similar to an SU UMa-type super outburst in profile. We show both events in \autoref{fig:LNUmaTESSevents}.

\begin{figure}
	\includegraphics[width=\columnwidth]{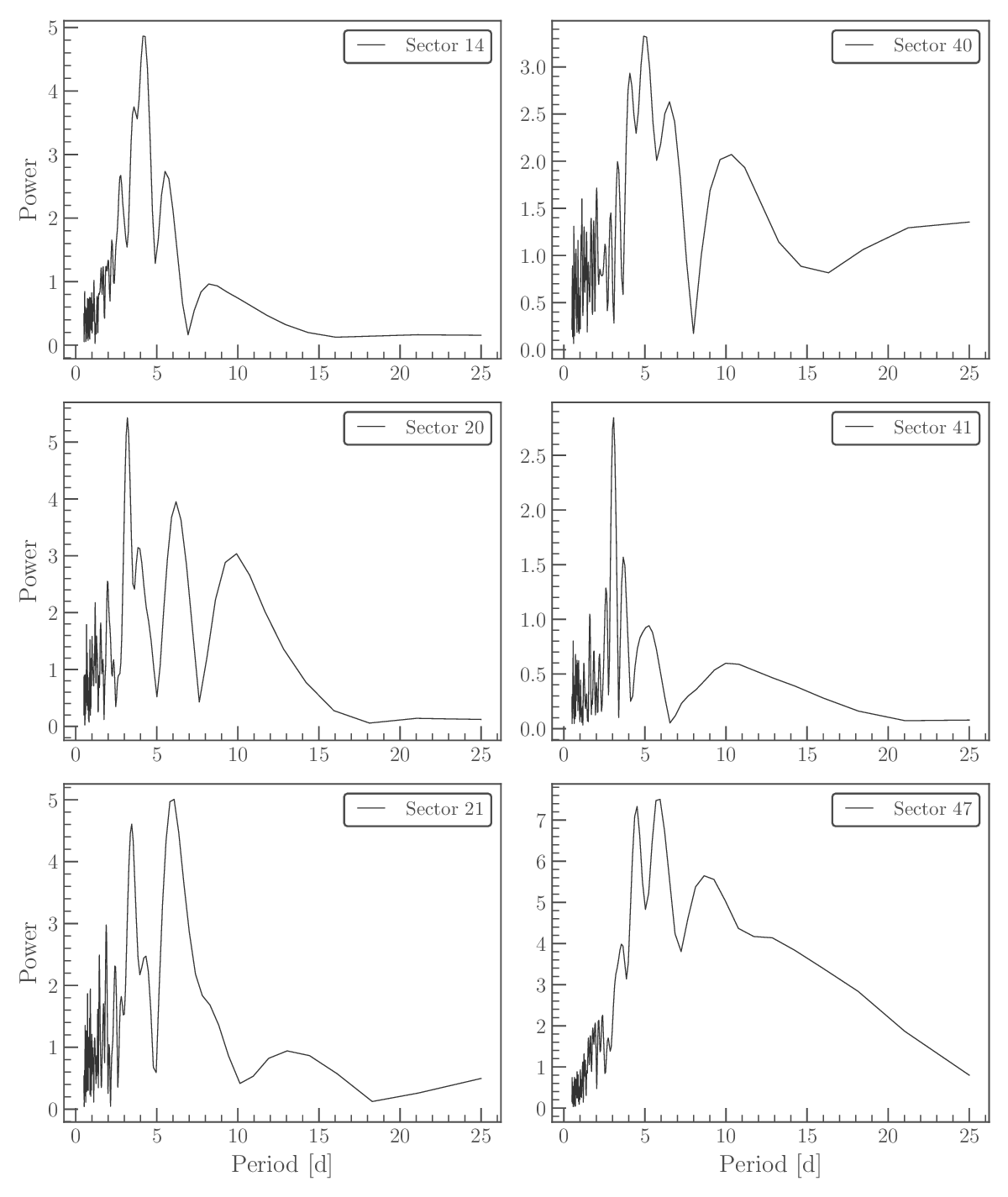}
	\caption[TESS periodograms of LN UMa]{The Lomb-Scargle periodograms of each of the TESS sectors of LN Uma observations showing a variable period between 3 and 6 days.}
	\label{fig:LNUMaTESSPer}
\end{figure}

In addition to this, each sector observed showed a periodic brightness variation. Period searching each of the sectors revealed a period, which varied by sector, of between 3 and 6 days as shown in \autoref{fig:LNUMaTESSPer}. In many sectors this variation contributes the most power (or a close second) to the periodogram, more than the orbital period. In the sectors which we show in \autoref{fig:LNUmaTESSevents} this variation can be clearly seen.

\subsubsection{V794 Aql}
High state observations of V794 Aql by \textit{ZTF} reveal a substantial variation of up to $\pm1$ mag centred upon $\rm mag\sim15$, previously highlighted by \citet{2014AJ....147...10H}. This variation appears in both of the observed bands. Period searching on this high state variation revealed signal between 17 and 19~d. {These are however, unlikely to be similar to the other period features we present in this work (\textsection\ref{sec:FeatPeriodic}). We consider that this feature is likely associated with the accretion disc due to its absence from observations of the low state. Indeed \citep{2014AJ....147...10H} identified these features as ``small outbursts'' which they saw to reoccur on a period of $\sim$15--50 days, which is entirely consistent with the data we present here. A consequence of this interpretation is that we are required to reassess the value attributed to the high state brightness, setting it equal to the quiescent brightness between these features ($\rm mag\sim16$--17), which in turn affects any amplitude calculated (which is reflected in \autoref{tab:overview})}

We also see two transitions into and out of relatively short lived low states, with both at minimum brightness for $\sim$30~d. It is not clear if the system truly completes the transition out of the first low state, however, as though the transitions peak close to the mean high state brightness the large brightness variation does not resume. As such it is possible that V794 Aql does reach high state again only to near immediately begin another transition to a low state but alternatively it is also possible that this an abortive transition out of the low state which fails to reach the high state. The measured transition speeds are consistent with those seen in other systems lying approximately in the middle of the distribution of \textit{e}-folding values which we measure. After the low state events the previously seen high state behaviour reasserts itself, however the extent of the depth of the variation increases by $\rm mag\sim1$ before gradually narrowing back to its initial behaviour. Similar to the behaviour which we report in BZ Cam, though to a greater extent, V749 Aql becomes redder at lower brightness.

\subsubsection{BZ Cam}
There appears to be a low state early on in the \textit{ZTF} observations which unfortunately falls during a visibility gap in observations; the early time data is insufficient for us to ascribe it as either part of the high state or early 
transition. However, observations after the visibility gap indicates the end of a transition from a low state. Other studies have identified low states occurring rarely where BZ Cam drops to a relatively shallow brightness of $V=14.3\rm\ mag$ \citep{1988PASP..100.1522G,2001A&A...376.1031G}, which the \textit{ZTF} observations 
are entirely consistent with. The transition we observe is by far the slowest we measure; this in part is due to the lack of data covering the entire transition although it is likely that its apparent slowness is a feature that a more complete data set would confirm. The colour distribution in BZ Cam has a relatively narrow spread, however with a clear trend to become redder at lower brightness.

\begin{figure}
    \includegraphics[width=\columnwidth]{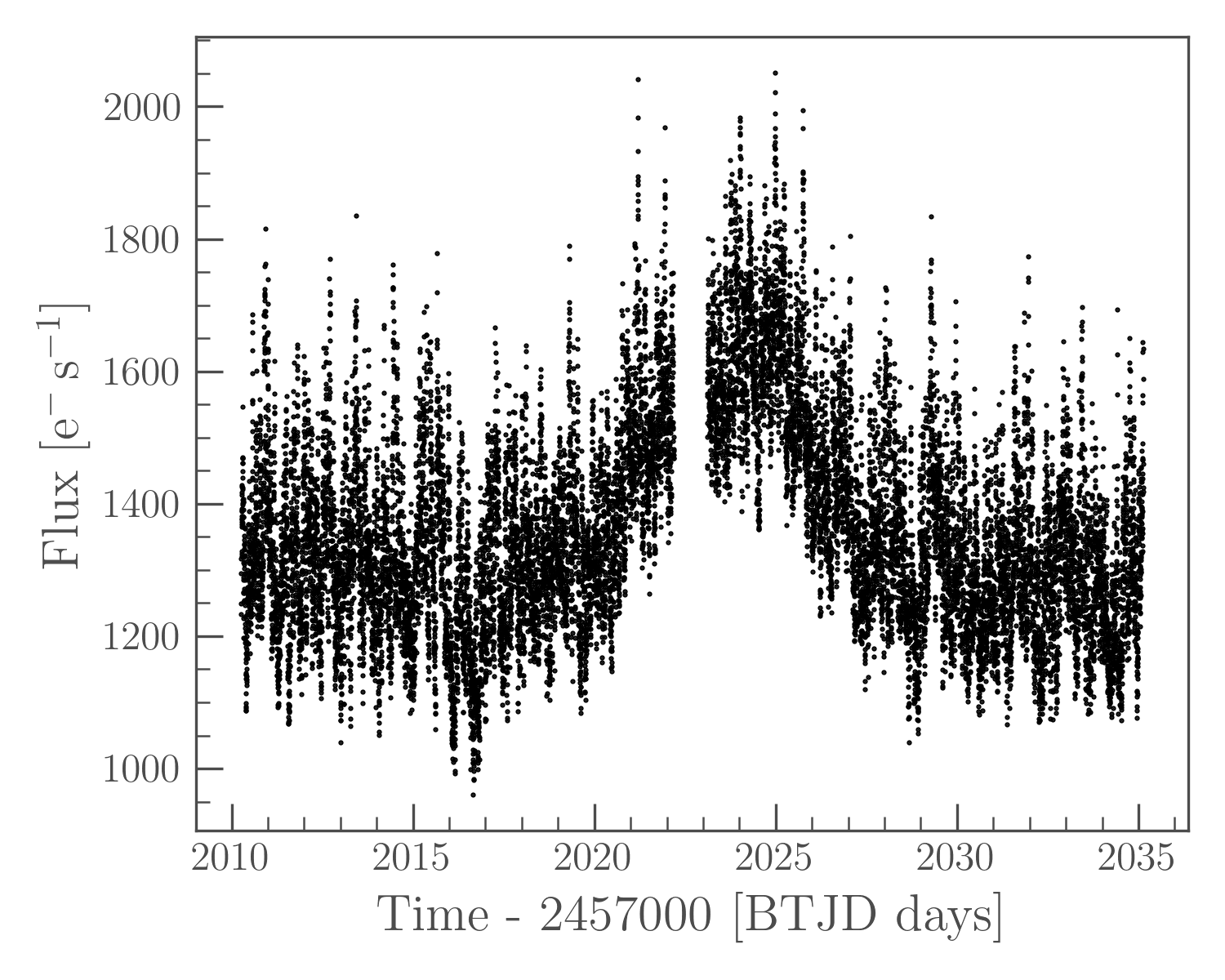}
    \caption{Extract of the \textit{TESS} lightcurve showing the sector 26 event BZ Cam}
    \label{fig:BZCamTESSevents}
\end{figure}

The \textit{TESS} observations of BZ Cam cover the high state, the majority of which show steady state behaviour. In sector 26, however, we see a $\sim$80\% increase in brightness in an event lasting approximately 10 days, shown in \autoref{fig:BZCamTESSevents}. The event, which takes on symmetric profile, is not proceeded by a dip in brightness (c.f. \textsection\ref{sec:lnumaZTF}). Unfortunately it falls in a \textit{ZTF} visibility gap so we have no complimentary data with which to compare this event, but we believe that this is  perhaps a short lived increase in mass transfer rate. 

\subsubsection{ES Dra}\label{sec:ESDra}

ES Dra is seen in the first $\sim$400~d of \textit{ZTF} data at a high state of $\rm mag\sim15$ consistent with previous high state observations \citep{2012NewA...17..108R} who identified the system as a Z Cam type. Despite this, \citet{2022arXiv220500632K} identified the system as both a Z Cam and VY Scl possessing low states and transitions in which no outbursts occur. Following the initial high state we see a transition to a depth of $\rm mag\sim17$, which is the known brightness of the low state in this system \citep{2022arXiv220500632K}. Following this singular transition, the \textit{ZTF} data coverage decreases dramatically making it difficult to describe the behaviour of lightcurve with degree of certainty. In order to overcome this challenge we supplemented the \textit{ZTF} data with data from \textit{ATLAS} \citep[Asteroid Terrestrial-impact Last Alert System;][]{2018PASP..130f4505T}. These additional data confirm the transition to a low state which is followed by a near immediate transition out of the low state, with the entire event being complete within $\sim$35~d. For the transition which we were able to measure, the data quality was not sufficient to yield a usable measurement, is at the fast end of the distribution of transitions measured. This may be associated with the relatively shallow low state similar to the transitions in V794 Aql. Following this, and similar to the findings presented by both \citeauthor{2012NewA...17..108R} and \citeauthor{2022arXiv220500632K}, there appears to be evidence of a dwarf nova outbursts \citep[c.f fig. 4][ see \autoref{fig:example} at $\rm T\approx 1600,1975$]{2012NewA...17..108R} consistent with Z Cam outbursting behaviour. Like some other systems which we have discussed in this work there appears to be a preference towards reddening in the low state.

The \textit{TESS} data for ES Dra, which is taken during the high state does not yield particularly insight to the behaviour of ES Dra, as it is spatially close to a bright solar type variable star. As a result much of the \textit{TESS} data is contaminated by the flux contribution this nearby star.

\subsubsection{KR Aur}\label{sec:kraurZTF}

Observations of KR Aur from \textit{ZTF} include two low states ($\rm mag \sim19$); the first starting before the commencement of the observations lasts for at least 500~d, while the second lasts $\sim$200~d and is caught almost in its entirety. Subsequent to each low state we also observe a high state ($\rm mag \sim14$), the first lasting $\sim$100~d and the second lasting at least 400~d extending past the end of observations. The transitions between the states show speeds typical for the those presented in this work, falling in the middle of the distribution; there is however a marked difference between the transition into the low state which is much faster than those where it is returning to the high state.

\begin{figure}
    \includegraphics[width=\columnwidth]{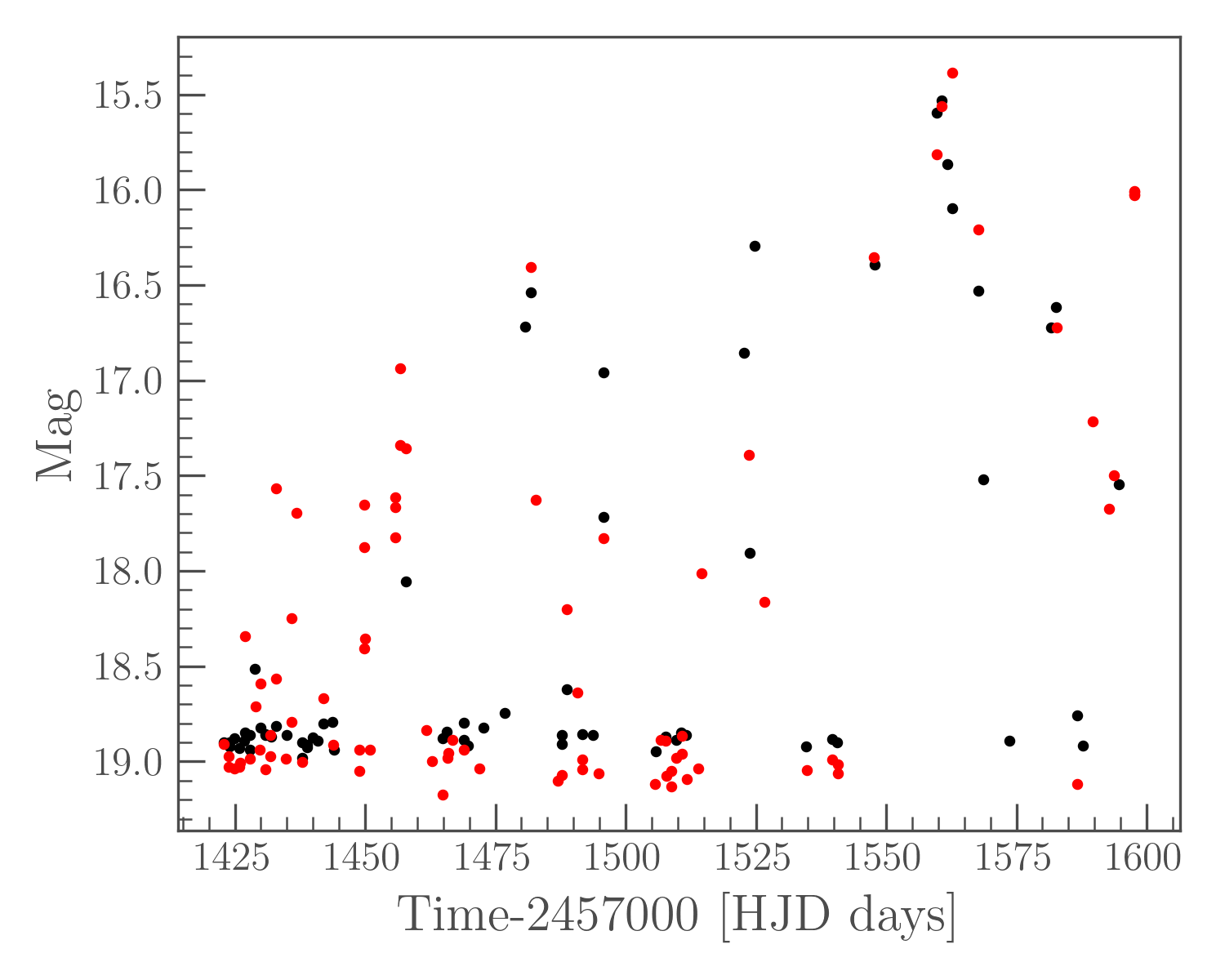}
    \caption{Extract of the ZTF \textit{g} and \textit{r} band, shown in black and red respectively, of KR Aur showing the successive ``outburst'' events approaching the event of the low state.}
    \label{fig:KREvents}
\end{figure}

Approaching the end of each of the low states we observe a number of ``outbursts'', an example of which we show in \autoref{fig:KREvents}, the amplitude of each being greater than that which preceded it. The initial ``outbursts'' have an amplitude of $\sim$1~mag with the final ones, immediately before the onset of transition, having an amplitude of $\sim$4.5~mag coming close to high state brightness. Although the sampling of the data makes it difficult to ascribe an exact duration to these events they appear to increase from $\sim$10~d to $\sim$30~d as the brightness increases. Due to the seasonal gap in data we are unable to confirm if these events continue up until the transition out of the first low state seen, though they are seen immediately before the transition out of the second low state. Unlike the events seen in LN UMa (\textsection\ref{sec:lnumaZTF}) these events are associated with a change in the colour index. The colour index increases dramatically during the ``outbursts'' with the brighter events attracting a greater colour index, e.g., they appear to be substantially more red than stable time observations, {which leaves these events at odds with typical DN outbursts and the predictions of the DIM \citep{2020A&A...636A...1H}}

%\textbf{\textit{Does this maybe suggest that the mass transfer rate doesn't change during these events and that although there is an enhanced accretion event, the material is expend such that the disc is depleted and the the effective temperature falls?}}

\subsubsection{V504 Cen}

\begin{figure*}
	\includegraphics[width=\textwidth]{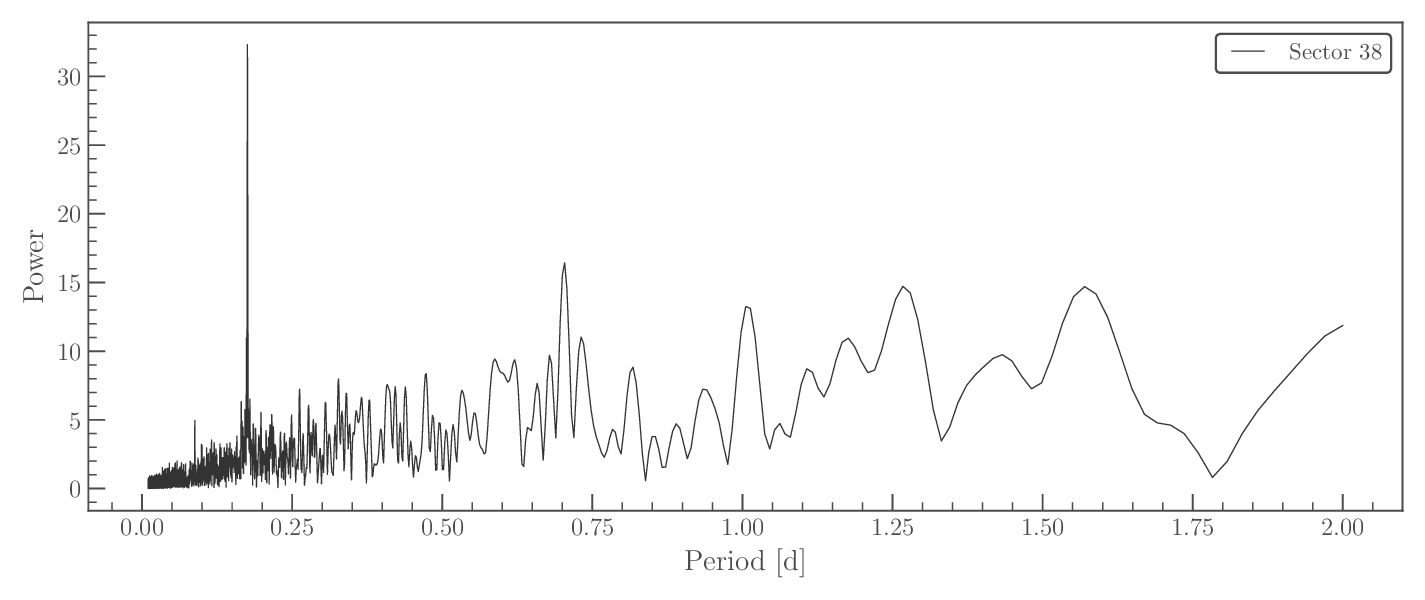}
	\caption[TESS periodogram of V504 Cen]{The Lomb-Scargle periodogram of the TESS observations of V504 Cen showing mostly strongly the orbital period followed by the signal at 4 times longer than the orbital period.}
	\label{fig:V504CenTESSPer}
\end{figure*}

{The 20~s \textit{TESS} photometry of V504 Cen reveals, as shown in \autoref{fig:V504CenTESSPer}, in addition to the orbital period, a peak in the power spectrum at four times the orbital period. This is consistent with the findings of \citet{2022MNRAS.514.4718B} who considered 120~s \textit{TESS} photometry of V504 Cen. That work however was unable to attribute any physical meaning to this long period, discounting the possibility of negative superhumps due to the significant deviation from the orbital period of the system. We agree with the assessment of \citeauthor{2022MNRAS.514.4718B} that this period is far too long relative to the orbital period to be a superhump, however the exact physical origin, of this period is not at all obvious, though the absence of lower harmonics of the orbital period would appear eliminate that as the source of this feature. We suspect, if physical, that it may be associated with some form of disc accretion. Using the relation developed by \citet[Eqn. 1,][]{2013MNRAS.435..707A}, which relates such a superorbital period to a negative superhump period, does not direct us to even a marginal peak in the power spectra indicating that indeed no superhump is present in this system.}

\subsubsection{MP Gem}
Recently identified as a VY Scl-type system by \citet{2021arXiv211107241K}, using the same \textit{ZTF} data we present here, prior to this its classification was unknown. The \textit{ZTF} data shows a single deep low state through the middle of 2018; the first observed since 1944 \citep{1963AN....287..169H}. The \textit{ZTF} observations commence with the transition into the low state well advanced, additionally, a visibility gap arises during the low state which makes it difficult to determine the exact duration of the state and impossible to determine its true depth. We were able to mitigate this issue by supplementing the \textit{ZTF} data with data from AAVSO\footnote{The American Association of Variable Star Observers: \href{https://www.aavso.org}{aavso.org}}. This helps us to constrain the duration of the low state to $\sim$300~d. Due to the visibility gap it is difficult to interpret the colour data with respect to the low state; only two data points are present from the low state which makes conclusions impossible to draw and leaves open the possibility that one, or both, are outliers and as such further observation of any future low state would be needed to confirm the colour behaviour.

\section{Overview of photometric features} 
From our study we have identified three features that appear in VY Scl systems: low states and their associated transitions, enhanced mass transfer events, and (quasi) periodic variations. We now address these in turn.

\subsection{States and Transitions} 
Transitions into low states are the defining feature of VY Scl binaries---they are not, however, all made equal. From the initial sample of 23 systems which we considered (see \autoref{tab:original}), we identified only 7 which featured low state behaviour in the coverage provided by \textit{TESS} and/or \textit{ZTF}. In addition, we also see substantial differences in how this behaviour relates to the recurrence time and duration which vary substantially. \autoref{fig:colourDisto} shows the distributions of colours and brightness states in the \textit{ZTF} systems. Restricting ourselves to just the brightness states initially, we see for example that both KR Aur and RX J2338+431 have a similar sampling and show at least one low state; however, they have strikingly different brightness distributions. KR Aur has a roughly bimodal distribution spending approximately equal amounts ($\sim$35\% per \autoref{tab:overview}) of time in both states, with the remaining time spent at intermediate brightnesses corresponding to transitions or ``outbursts". RX J2338+431 in contrast strongly favours high state brightness with only a very small number of days at low state or transitions. 

This pattern of variability in low state behaviour highlighted here is repeated across the sample of VY Scl systems which we present in this work, and is immediately apparent visually from \autoref{fig:example} and is quantitatively supported by \autoref{tab:overview} where we see a spread in the fraction spent in the observed high state values and depth of the low states. It is, however, unclear what its physical origin of this variation is, with no apparent correlation to {orbital period (which is often correlated to a variety of system parameters in CVs)}.

In \autoref{tab:eFolding} we show the \textit{e}-folding time for each of the transitions which we have observed. This shows that the transition speeds generally fall within a broad range, 16.7~d--33.5~d, with a few outside this range. The distribution of the speed of transition appears to have no association with the measured orbital period of the system; there is a weak correlation between the depth of the transition and its speed with deeper transitions being slower in general---although this discounts BZ Cam which has a very slow but apparently shallow transition. {This could be understood in the context of deeper transitions arising in systems where the disc is a larger contributor to the overall brightness and therefore physically larger. In this circumstance a physically larger disc would likely shrink or dissipate more slowly than a smaller disk.} We also see that the fastest transitions occur in those systems which almost immediately transition out of their low state instead of remaining in a sustained low state, {this suggests that some ``inertia'' exists in the mechanism which suppresses mass accretion -- the longer it is in place the more established it becomes.}

Within individual systems the speed of transitions are more consistent; however, we observe that there is an appreciable difference between the speeds into low state and out of low state transitions. Continuing in the spirit of these systems behaving differently to each other, we see that two of the systems show faster transition into the low state, whereas three systems are faster when leaving the low state. There is no relationship between such systems and the orbital period of the system, similarly the amplitude of the transitions appears to show no correlation with these behaviours. As there appears to be no theoretical predictions surrounding the speed of the transitions it is difficult to ascertain if there is a genuine physical process at work due to the relatively small numbers considered in this work. Despite this, given that we observe in those systems which have more than a single pair of transitions a consistency in behaviour, suggesting it is probable that this is not an artefact of a selection bias but a real feature.

Turning to the variation in the colour of the systems between high and low states we see that once again there is no one single behaviour observed in these systems, with differences between both high and low states seen across multiple systems. We see in at least three systems a preferential reddening at lower brightness, whilst the remaining appear to show no difference in colour between the different states. In addition to this we see that for \textit{all} systems during their respective high states that there is a positive correlation between brightness and blueness; this is entirely inline with expectation as within a high state brightness variations associated with small changes in the mass transfer rate would be expected and slight increases would increase both the brightness and the temperature of the system, hence the observed blueness relation.

\subsection{Periodic behaviours}\label{sec:FeatPeriodic}

Three of the systems which we have discussed appear to show periodic variations in their lightcurve which are identified by the high time resolution made available to us by \textit{TESS}. A number of periodic behaviours have been identified previously in CVs in addition to the orbital and spin periods {(\autoref{tab:overview})} of the systems. These include positive and negative superhumps, superorbital periods, and the associated beats and harmonics that arise from the interaction of these periods. Superhumps have frequencies within a few percent of the orbital period: physically negative superhumps are associated with a warped accretion disc which has a retrograde precession in the inertial frame \citep{2015ApJ...803...55T}, whilst positive superhumps are associated with a resonant tidal stress in the accretion disc induced by the secondary star driving a ``flexing'' disc oscillation in the orbital plane and a positive precession in the inertial frame \citep{1991MNRAS.249...25W,2000ApJ...535L..39W}. The periods of precession (retrograde and/or prograde) are called superorbital periods, and can make an additional contribution to the power spectrum of a given system over and above what might be expected from the interaction of the superhump and orbital period \citep{2013MNRAS.435..707A}.

The \textit{TESS} lightcurve of LN UMa varies on a substantially longer timescale than any of the other sytems that we have presented, with a frequency well below $1\rm\ d^{-1}$ which varies depending on the sector considered; giving periods between 3 and 6 d. Placing this even further at odds with the other systems this very low frequency variation is often the dominant frequency in the lightcurve such that it can be easily picked out by eye in the lightcurve (see Figures \ref{fig:LNUmaTESSevents} and \ref{fig:LNUMaTESSPer}). If this is a superorbital period associated with the disc precession then it indicates both a very slow precession which makes a substantial contribution to the flux of the system.

\section{Discussion}  
\label{sec:discussion}

\subsection{Light curve morphology dichotomy}
\label{sec:di-morphology}

The high mass transfer rate during their high state and the lack of outbursts in the low state are common properties of all VY Scl systems \citep{warner2003cataclysmic}. We argue that these common features indicate the presence, or absence, of certain mechanisms intrinsic to these CVs. However, there are also observational features found in some but not all VY Scl systems. For instance, from the visual inspection of the light curves shown in \autoref{fig:example}, we may broadly divide them into two subgroups {based upon their brightness variations\footnote{We define this by setting DM as the difference between the magnitudes of the low and the high  state and dm as the variation of the magnitude of the source during their high state. For MV Lyr, DM = 5 and dm < 1; and for V794 Aql DM = 4 and dm = 2. Thus, dm/DM < 0.2 for MV Lyr and dm/DM ~ 0.5. Therefore, MV Lyr has  relatively weak fluctuations in the high state and V794 Aql has substantial fluctuations in the high state}}: (i) systems with a relatively weak brightness fluctuations in their light curves in high states (Group A systems), and (ii) systems with substantial flickering in the light curves in high states (Group B systems). With this in mind, we now attempt to construct a framework, in terms of mass transfer, accretion in interacting binaries and the feedback reaction to mass transfer processes and the orbital dynamics, to account for the phenomena associated with the high-low states observed in the VY Scl binaries in a holistic matters.  

%From the morphology of the light curves, 
%  we divide the systems into two groups 
%(see caption of \autoref{fig:example}  
%  for the designation of Group A and Group B systems). 
The most noticeable feature of the Group A systems is that there is a brightness ceiling during the high states. Moreover, the low states are episodic, where the brightness either drops dramatically from the ceiling brightness, e.g. in RX~J2338+431 and ES~Dra,  or the source becoming substantially dimmer for a reasonable long duration, e.g. in MV~Lyr, KR~Aur and MP~Gem.   The light curves of the Group B systems are complex and structured and appear not to have a brightness ceiling during the high state. They, however, exhibit flare-like fluctuations in the high state, e.g. in LN UMa. Although the entry and the exit of the low states in some systems, e.g. V794~Aql, can be identified, the brightness variations during the state transitions in the other systems are not easily characterised.  

\subsection{Scenarios on high-low states and state transitions}
\label{sec:States}  

%VY Scl binaries are observed 
%    to exhibit episodes of high and low brightness, 
%    and these episodes are 
%    referred to as high and low states respectively.  
The high and low states in CVs are often attributed to the high and low rates of mass transfer from the mass donating star to the white dwarf. This association may be appropriate for magnetic CVs, such as Polars and Intermediate Polars \citep{2022MNRAS.516.3144D,2024ApJ...965...96M}, but problematic when the binary has a fully developed accretion disc.  

In CVs, the secular mass transfer rate averaged over episodes of high and low states, are determined by the efficiency of angular momentum loss \cite[see e.g.][]{Verbunt1981A&A}. The high mass transfer rates in VY Scl binaries, relative to other CVs with similar orbital parameters, therefore requires additional mechanisms that increase the loss or transfer of angular momentum from the orbit to the component stars. 

Observations have shown that VY Scl binaries are more often in the high state than in the low state. Moreover, their high brightness  in the high states cannot be merely due to stability in the accretion disc, implying that the mass transfer also occurs at a very high rate in the VY Scl binaries \citep[see][]{Zemko2014MNRAS}. The two pieces of observational evidence must be accounted for when building a model to explain secular behaviour of VY Scl binaries.  

While the high state of VY Scl binaries are often attributed with the accretion-disc instabilities \cite[see e.g.][for a review of disc instability]{Mineshige1993Ap&SS}, the state transitions are triggered by the change in the mass outflow from the companion star \cite[see][]{King1998ApJ}. Changes in the mass transfer in CVs can be caused by activity intrinsic to the companion star, such as stellar magnetism, or certain external processes, such as certain interactions between the companion star and the WD. The companion star in CVs are low-mass stars which can show evidence of strong magnetospheric activity \citep[see][]{Berdyugina2005LRSP}. The migration of multiple magnetic star-spots or a giant star-spot into the region near the L1 point  will alter the outflow of material  from the donor secondary star \citep{1994ApJ...427..956L} which is filling its critical Roche surface. The magnetic fields carried by the star-spot(s) into the L1 region, depending on the orientation, can create a barrier blocking the flow of material through the L1 point into the WD's Roche lobe. Mass transfer resumes when the star-spot(s) migrate out the L1 region. This process is magneto-hydrodynamical (MHD), not directly associated with the evolutionary state companion star (which is of the longer Kelvin-Hemholtz time scale), the orbital dynamics of the binary, or the physical condition of the accreting WD. 

Despite being appealing, the star-spot scenario does not explain the prolonged low state as observed in some systems, e.g. KR Aur.  
Inspection of the light curves of the Group A systems (see \textsection\ref{sec:di-morphology}) also reveals that their brightness does not show large-amplitude brightness flickering, when they enter the low state. Plasma and MHD processes generally exhibit instabilities, which can trigger flaring and eruptive behaviour, which would be seen as flickering. Therefore, the mass transfer in these VY Scl binaries is either absent or relatively steady during the low state. This also hints at the presence of some secular processes, at least partially responsible for, driving the Group A systems into the low state, and these driving processes are relatively steady, evolving over timescales significantly longer than the duration of the high-low state and their transitions. Migrations of magnetic star-spots of the companion star are on relatively short timescales and therefore are unlikely to drive the Group A systems from a high state or confine them in a prolonged low state.  
  
As opposed to the star-spot scenario, the high state is now not necessarily the default state, and the low state is not simply a temporary disruption of the mass transfer. This puts the high and low states in a similar footing in the modelling, where the average mass transfer is driven by the secular orbital evolution, with magnetic braking and/or gravitational radiation as the angular momentum loss mechanisms \citep[see e.g.][]{Verbunt1981A&A}. {It should be noted however, the understanding of the drivers of secular orbital evolution are not completely understood as there is strong irradiation of the M star which could give rise to non-linear feedback \citep[see][]{Wu1995PASA}}

%Also, different processes are required 
%  for triggering the state transitions  
%  and for the maintenance 
%  of the high and the low states.   
%  
%The orbital periods of VY Scl binaries 
%  are between 3 to 4.5~hr, and 
%  a semi-detached binary with such periods 
%  would have a companion star of mass 
%  about $0.3-0.4\;\!{\rm M}_\odot$. 
%A star will become fully convective  
%   when the mass drops below 
%   around $0.3-0.35~{\rm M}_\odot$ 
%  \citep{MacDonald2018MNRAS,Baraffe2018A&A}.   
%The companion stars in VY Scl binaries 
%  therefore either posses  
%  a very small radiative core or 
%  have become fully convective.   
%The latter involves a sequence of processes 
%  operating over the entire binary  
%  in order to trigger the state transitions 
%  and to sustain the high and the low states.

%The scenarios 
% proposed by \citet{1999MNRAS.305..225L}, 
% \citet{2002A&A...394..231H}.
% and  \citet{Wu1995PASA} 
% are respective examples in this subclass. 

%\noindent  
%\citep{Page2014} \\ 
%\citep{Zemko2014MNRAS} \\ 
%\citep{Sion2009JPhCS} \\ 
%\citep{Hamilton2008PASP} \\ 
%\citep{Greiner1998A&A} \\ 
%\citep{2018A&A...617A..16S}

\subsection{A more holistic consideration}

\citet{1999MNRAS.305..225L} and \citet{2002A&A...394..231H} present explanations as to the absence of disc instability in VY Scl systems, but are markedly different, relying on entirely different physics to arrive at their conclusions ---\citeauthor{1999MNRAS.305..225L} uses disc irradiation whereas \citeauthor{2002A&A...394..231H} relies upon weak WD magnetism eliminating the accretion disc. 

Due to the weakness of the proposed magnetic fields required by \citeauthor{2002A&A...394..231H}'s model and the data that we have considered in this work, we cannot address this theory directly; had we had \textit{TESS} data of low states it may have been possible to confirm the presence of a disc by identifying periodic features originating from a disc. Despite this it remains part of our consideration throughout this work. We can however address the disc irradiation model with the present data.

The disc irradiation model states that during the low state the WD is sufficiently hot to irradiate the inner accretion disc such that it can have a temperature which would ordinarily be associated with the stable high state. As the system is substantially fainter in this configuration, we would anticipate that the relative contribution of the irradiated disc to the total brightness would increase. Under this scenario, the irradiated disc, having remained hot, should produce a greater blue excess than would be expected for a system in a low state, which may manifest as simply not seeing an increase in red excess. Consequently, we believe that those VY Scl stars which do not become redder during their low state are showing evidence of an irradiated inner disc.

As noted previously the colour behaviour during low states is not a universal behaviour, with many of the systems becoming redder during the low state. {Continuing with the hypothesis that colour evolution can be used as a tracer of an irradiated disc, this would imply that these systems lack such a disc,} and as noted by \citet{2005astro.ph..6382H} some cases exist (including MV Lyr) where the weak magnetic field model successfully holds at the expense of the disc irradiation model. Without being able to probe the the weak magnetic field model with the data presented here we cannot comment directly on its validity, however comparing our data with that discussed in \citet{2005astro.ph..6382H} we see no grounds for disagreement. With this in mind we believe that \textit{both} models are valid and apply to systems according to the colour behaviour which they exhibit. We further believe that this is the case because it seems unlikely that all of the WDs across the VY Scl binary population are magnetic, first because we (and others) have seen evidence for ``outburst'' like behaviour most plausibly originating in a disc and secondly because WD population studies seeking magnetism below the threshold required for magnetic disc disruption have identified a substantial fraction of the population are not magnetic \citep{2021MNRAS.507.5902B}.

The other challenge of the behaviour in VY Scl binaries is understanding why during the transitions we do not observe outbursts, as these generally take longer than the viscous time. Whilst the magnetic model is capable of eliminating the disc for a sufficient duration to explain the absence of outbursts, the disc irradiation model has more difficulty. As the disc grows with increasing mass transfer rate during the transition it becomes increasing difficult for irradiation to keep the disc hot enough to prevent instability developing (the reverse of this issue applies during transitions into low states). If we consider the transitions of those systems which we believe to be disc irradiation systems with those likely governed by magnetism we see that there is no discernible relationship between the colour behaviour and the speed, $\tau$, or the duration of the transitions. With that being the case we do find that the shortest mean duration for transitions is in those systems which are preferentially red in the low state however this does not extend to all such systems. Consequently we do not believe that this is can offer any explanation as to the transition behaviour which we observe.

As we have already highlighted, there is a wide range of transition speeds and low state durations present in the VY Scl binaries we have considered here. We see in those systems with the shortest low states, e.g., RX~J2338+432 and ES Dra, that these systems have the fastest transitions which we measured. Similarly, we see in those systems with much longer low states, e.g., MV Lyr and MP Gem; that these systems generally have far slower transitions. Although there are some exceptions to this, for example KR Aur where other factors may be involved (see \textsection\ref{sec:EnhancedMass}), there does appear to be a correlation between the speed of transition, which may be connected to the aforementioned weak correlation between transition speed and state depth as longer low states tend to be be deeper. This indicates that there exists a difference in the mechanism of the transitions in shorter shallower low states from those longer deeper low states.

%\textbf{To address...
%\begin{itemize}
%   \item What does this maybe tell us about the mechanism controlling these states? 
%    \textcolor{blue}{[KW] Some interplay between 
%     the mass transfer instability.}
    %\item Is the speed of transition effected by the longer term behaviour of the system? eg. time since last transition etc. 
   % \textcolor{blue}{[KW] It does not look like it when I checked the light curves shown in the paper. }
%\end{itemize}
%}

\subsection{Enhanced Mass Transfer}
\label{sec:EnhancedMass}

Although the absence of outbursts is a defining feature of VY Scl binaries we present evidence of ``outbursts'' or enhanced mass transfer events. This would not be the first time such events have been seen in VY Scl binaries with previous observations of magnetically gated outbursts \citep{2017Natur.552..210S} and ``stunted outbursts'' \citep{2004AJ....128.1279H}. The events which we observe, however, are diverse in nature and with the exception of V794 Aql, do not appear to have been previously identified and seem to be different from those which been remarked upon before.

Like the stunted outbursts seen in V794 Aql, the events which we see in LN UMa, have a quasi-periodic nature, with a similar duration and recurrence time, which indicates that it is appropriate to identify them as ``stunted outbursts''. The LN UMa events however have some key differences, firstly they are not ubiquitous to the high state; indicating that the instability which causes them is not permanent. Furthermore, unlike V794 Aql the events in LN UMa are not symmetric and always commence with a fall in brightness before entering a rise phase. This indicates that mass transfer is (partially) disrupted for a short time before increasing above the normal high state rate. This suggests that when mass transfer is disrupted the mass that ordinarily flows during this time builds up at some boundary before being released resulting at an enhanced rate and consequently brightness. It is likely that the physical origin of these events are linked through mass transfer instability, with LN UMa having an additional contribution from the magnetic field causing the mass to be held at some boundary \citep[c.f. magnetic gating;][]{2010MNRAS.406.1208D,2012MNRAS.420..416D}.
%\textbf{To address... What could cause this...
%\begin{itemize}
%    \item Magnetic gating? See DW Cnc outbursts for example?
%    \item Transient sunspot passing L1? Seems unlikely given the repeated nature of the event
%    \item \textcolor{blue}{[KW] Two types. One with mass transfer instability, another with both mass transfer instability and 
%    B interplay. I need some quick calculations on the time scales 
%    .... }
%\end{itemize}
%}

The events we present in KR Aur are particularly interesting, with parameters which appear to be directly tied to the imminence of a transition to the high state. These ``outbursts'' get progressively greater in amplitude {(1.5-4.7 mag)} and redder {(0.15-0.70)} in colour as the system nears transition. This implies that physically these events involve material that is locally cooler than the rest of system. If, as we suggest in \textsection\ref{sec:States}, KR Aur is a system which has a disc it is possible that these events indicate a increase in the mass transfer rate from the L1 region which results in only part of the accretion disc being inflated in size and brightness before fading again as material dissipates returning to the quiescent brightness. As the mass transfer rate continues to increase so does the scale of the events until the mass transfer rate is sufficient to trigger an all disc effect --- transition back to the high state. Physically, this may be an extreme case of the flaring observed in the low state of the group A systems (see \textsection\ref{sec:dis-phen}), which induces some feedback mechanism to inflate the buffer annulus.

The final remarkable event which we have observed is the relatively long outburst observed in BZ Cam. Lasting $\sim$10~d and with a relatively small increase in flux this is indicative of a stunted outburst. With the \textit{TESS} coverage which is available we would ordinarily expect to identify more that one such event, however without consecutive sectors of observation and allowing for a low recurrence time there are many plausible recurrence timescales where we would miss subsequent events. That explanation is not satisfactory to explain the apparent absence in \textit{ZTF} data. It is possible that these events are hidden in the stochastic variation of the high state lightcurve {(which is at the upper range of ZTF's capability)}, although the flux increase seen in the \textit{TESS} data indicates that this may be unlikely. Revisiting the lightcurve, there are occasions (e.g., $\rm HJD\approx2458500$ and $\rm HJD \approx2459150$) where the brightness exceeds the local neighbouring values---it is possible that these are further events that we simply cannot resolve. Unfortunately these events occur during gaps in the \textit{TESS} coverage; however if we assume that these are similar events then combining with the resolved event then we see a recurrence timescale of $\gtrsim$100~d. {Subsequent AAVSO data would appear to support this, with at least 2 similar events present (e.g., $\rm JD\approx2459000$ and $\rm JD \approx2459630$).} This most likely cannot be definitively addressed without more long term observations which are suited to detecting such brightness changes in relatively bright sources.

\subsection{A phenomenological model for the brightness variations and brightness ceiling in the high state}\label{sec:dis-phen}
  
The apparent ceiling of the brightness of the Group A system implies their maximum brightness is determined by a set of variables that are relatively stable and do not vary over short timescales. One possibility is the formation of an optically thick annulus boundary layer with a small width attaching to the white dwarf. The outer boundary of the opaque boundary layer contacts a buffer annulus, which acts as a mass reservoir and regulates the inflow of the material into the inner opaque boundary layer. The maximum brightness is therefore determined by the temperature and the size of ``photosphere" corresponding to this boundary layer, which are determined by the mass and radius of the accreting WD. The high state therefore corresponds to the situation where the buffer annulus is filled and the low state corresponds to the situation where the buffer annulus is empty. Moreover, the radiative force must play an important role, and the flows there are thermally stable (absence of cooling induced instability), weakly magnetic (avoiding eruptive magnetic reconnection) and non-convective (allowing laminar circular flow).  

In this scenario flaring can occur during low states, where the buffer is almost empty and the optically thick annulus boundary layer cannot be formed. The magnetic activity and hydrodynamic instability in the resident flow can trigger flares. In a certain way, the structure of the disk is analogous to a Solar-like star but in 2D, where an radiative inner region is bounded by a convective outer region. 

The Group B systems are those where the optically thick radiative dominate annulus boundary layer cannot form. A possibility is that the WDs in the Group B systems have a strong magnetic field, which have non-uniform strengths and orientation distributions over the WD surface. The magnetic-field causes instability, disrupts the flow, and the optically thick annulus boundary layer; hence the buffer annulus outside it cannot be developed. The high low states of the Group B systems are triggered by the variations in the secular mass transfer, and the flaring simply reflects the instability in the flow triggered by the magnetic field in the accretion disk which also interacts with the magnetic field of the WD. This scenario can also easily accommodate the observations that some VY Scl binaries show behaviour similar to that of super-soft source \citep[see][]{Greiner2010} and that a strong wind can be launched from the disk of VY Scl binaries \citep{Inight2022}. 

The remaining questions are now how to maintain the high state of VY Scl in Group A and how to model the respective accretion disks in the Group A and Group B systems. One scenario to explain systems which can sustain high-mass transfer rate (during the high states) above the average mass-transfer rate set by orbital evolution due to angular momentum loss is irradiative heating of the mass donor star \citep[see e.g.][]{Wu1995PASA}. This is easier to achieve in the Group A systems as their high states tend to have a stable ceiling luminosity from the radiative dominate flow near the white dwarf. The answer to the latter question is less trivial. It will require a more sophisticated accretion disk model that can take care of the complex interaction between magnetic field interaction within the accretion disk and between the white dwarf and the accretion disk together with relevant radiative hydrodynamics processes self-consistently. 
  
\subsection{Remarks on the Periodic Behaviours}

We see three different manifestations of notable periodic behaviour in the systems we have considered in this work; a 1.1~d period in TT Ari, a $\sim4P_{\rm orb}$ period in V504 Cen and a $\sim$3--6~d period in LN UMa. {The variety of features observed indicates a diversity in the physical properties of the systems.} Whilst the period in TT Ari can be explained reasonably simply (see \textsection\ref{sec:FeatPeriodic}) {such that may be seen in any CV with an accretion disc}, the others are more puzzling. As stated previously the periodic signal in V504 Cen cannot be either a superhump nor a superorbital period; however {it is clearly real as it exists in two separate data products.} As we have previously suggested, it may be some resonant accretion feature where the mass transfer rate varies on the timescale of the observed variation. A concerted observational effort will be required to establish the exact origin of this feature is and would benefit from study from a longer term view point in addition to that provided by \textit{TESS}.

Finally, we turn to LN UMa, which shows a very prominent low frequency oscillation that appears to change over relatively short time frames---although it is possible that we are detecting a harmonic or overtone of a single frequency. Nevertheless, without needing to ascribe an exact value to this period, we can say that this period almost certainly is real, but of unclear physical origin. If LN UMa is a system with high inclination\footnote{No value for the inclination of LN UMa is available, however as a member of the SW Sex subclass it is reasonable to assume that this is the case \citep{2005ASPC..330....3G}.}, it is possible that the inclination is generating some projection effect that affects the fraction of the accretion spot that can be seen, however we would expect that this would occur on a period that could be related reasonably easily to the orbital period. In order to fully understand this feature further, study will be necessary including an effort to determine if the period is fixed or not, and if not what the rate of period change $\dot P$ is.

{Unusual periods in CVs with physical origins that cannot immediately be identified is not an unheard of phenomenon. \citet{2022MNRAS.514.4718B} who studied the long term \textit{TESS} light-curves of 15 different CVs (including three systems we consider here) found several unusual but nonetheless real periodicities. For example, in AC Cnc they identify a 4.6 d period which they cannot identify an origin for. In V533 Her they identify variations in the superhump periods, in this case they ascribe the variability to interactions between different periods of different amplitudes. That work concludes that systems which have unusual periodic behaviours should be subject to concerted high cadence study with complimentary followup observations to understand system parameters which may be useful in understanding the physical origin of these periods.}

\section{Conclusions}
Although a relatively small sub-class of Nova-Like CVs, the fact that VY Scl systems show irregular drops in brightness of over 1 mag shows they are important sources with which to understand how the mass transfer rate in accreting binaries is regulated. Similar to other Nova-like systems, VY Scl systems do not show classical outbursts, however they do show low states. Additionally the DIM predicts outbursts should be seen during the low states {(if a disc is present)}, though the ``outbursts" which are present cannot be supported by the DIM. These outbursts have a range of behaviours including those of KR Aur which are repeated and become increasingly energetic throughout a low state. Further examples of stunted outbursts, {such as those we present here}, in VY Scl and other systems can help give further insight to their cause. 

Having presented long baseline and high time resolution observations of 10 known VY Scl systems, we have reviewed the models which set up to describe the behaviours of these systems and attempted to reconcile these with observations. We have presented a unified model to describe both the cause of the state transitions and the observed variability of VY Scl systems lightcurves and how this differs on a system to system basis.

The observations shown here have, at least in part, come from all-sky optical surveys. With the expansion of surveys, with some projects such as GOTO \citep{2022MNRAS.511.2405S,2022SPIE12182E..1YD} having telescopes in the northern and southern hemispheres, with the capacity to cover the entire sky every few days, the opportunity exists to greatly extend our length of coverage of VY Scl systems to gain a better long term view of their accretion behaviour. In particular, it will be possible to obtain multi-wavelength spectroscopy of sources which go into a new accretion state or which start a series of stunted outbursts. This will give essential information with which to improve accretion instability models to account for their behaviour.  

\section*{Acknowledgements}
This paper includes data collected by the \textit{TESS} mission. Funding for the \textit{TESS} mission is provided by the NASA's Science Mission Directorate.

It also includes \textit{ZTF} data  obtained with the Samuel Oschin Telescope 48-inch and the 60-inch Telescope at the Palomar Observatory as part of the Zwicky Transient Facility project. \textit{ZTF} is supported by the National Science Foundation under Grants No. AST-1440341 and AST-2034437 and a collaboration including current partners Caltech, IPAC, the Weizmann Institute for Science, the Oskar Klein Center at Stockholm University, the University of Maryland, Deutsches Elektronen-Synchrotron and Humboldt University, the TANGO Consortium of Taiwan, the University of Wisconsin at Milwaukee, Trinity College Dublin, Lawrence Livermore National Laboratories, IN2P3, University of Warwick, Ruhr University Bochum, Northwestern University and former partners the University of Washington, Los Alamos National Laboratories, and Lawrence Berkeley National Laboratories. Operations are conducted by COO, IPAC, and UW.

This work has also made use of data from the Asteroid Terrestrial-impact Last Alert System (ATLAS) project. The Asteroid Terrestrial-impact Last Alert System (ATLAS) project is primarily funded to search for near earth asteroids through NASA grants NN12AR55G, 80NSSC18K0284, and 80NSSC18K1575; byproducts of the NEO search include images and catalogs from the survey area. This work was partially funded by Kepler/K2 grant J1944/80NSSC19K0112 and HST GO-15889, and STFC grants ST/T000198/1 and ST/S006109/1. The ATLAS science products have been made possible through the contributions of the University of Hawaii Institute for Astronomy, the Queen’s University Belfast, the Space Telescope Science Institute, the South African Astronomical Observatory, and The Millennium Institute of Astrophysics (MAS), Chile.

This work was funded by UKRI grant (ST/T505936/1). For the purpose of open access, the authors have applied a creative commons attribution (CC BY) licence to any author accepted manuscript version arising. C. Duffy acknowledges STFC for the receipt of a postgraduate studentship.

Armagh Observatory \& Planetarium is core funded by the Northern Ireland Executive through the Department for Communities.

This work made use of Astropy:\footnote{http://www.astropy.org} a community-developed core Python package and an ecosystem of tools and resources for astronomy \citep{astropy:2013, astropy:2018, astropy:2022}.

This research made use of Lightkurve, a Python package for Kepler and \textit{TESS} data analysis.

We acknowledge with thanks the variable star observations from the AAVSO
International Database contributed by observers world-wide and used in this research. 
This work has made use of the NASA ADS. 

\section*{Data Availability}
{\it TESS} data is available from the Mikulski Archive for Space Telescopes (MAST), which can be accessed at \url{https://mast.stsci.edu/portal/Mashup/Clients/Mast/Portal.html}\\
\textit{ZTF} data is available via the NASA/IPAC Infrared Science Archive \url{https://irsa.ipac.caltech.edu/Missions/ztf.html}\\
ATLAS forced photometry data is available from the collaboration website \url{https://fallingstar-data.com/forcedphot/}

\bibliographystyle{mnras}
\bibliography{references}

\appendix
\section{Complete List of Considered Sources}
\begin{table*}
    \caption{Complete List of the sources considered for inclusion in this work, showing the survey(s) in which data was available at the point of analysis. Systems were excluded from analysis where they did not show state transitions or other features of interest.}\label{tab:original}
    \begin{tabular}{lccc}
    \hline
    Object Name & ZTF & TESS & Included for Analysis\\
    \hline
    V504 Cen &  $\times$& \checkmark & \checkmark \\
    VY Scl & $\times$ & $\times$ & $\times$ \\
    VZ Scl & $\times$ & $\times$ &  $\times$ \\
    V1082 Sgr & $\times$ & $\times$ &  $\times$\\
    V442 Oph & $\times$ & $\times$ &  $\times$\\
    V794 Aql & \checkmark & $\times$ & \checkmark \\
    LX Ser & \checkmark & \checkmark & $\times$ \\
    KR Aur & \checkmark & $\times$ & \checkmark \\
    RXJ2338+431 & \checkmark & \checkmark & \checkmark \\
    MV Lyr & \checkmark & \checkmark & \checkmark \\
    V751 Cyg & $\times$ & $\times$ & $\times$ \\
    BH Lyn & \checkmark & \checkmark & $\times$ \\
    V425 Cas & \checkmark & \checkmark & $\times$ \\
    LN UMa & \checkmark & \checkmark & \checkmark \\
    BZ Cam & \checkmark & \checkmark & \checkmark \\
    HS0506+7725 & $\times$ & $\times$ & $\times$ \\
    TT Ari & $\times$ & \checkmark & \checkmark \\
    DW UMa & $\times$ & $\times$ & $\times$ \\
    MASTER OTJ190519.41+301524.4 & $\times$ & $\times$ & $\times$\\
    RX J2338+431 & \checkmark & \checkmark & \checkmark \\
    MACHO 311.37557.169 & $\times$ & $\times$ & $\times$\\
    MP Gem & \checkmark & $\times$ & \checkmark \\
    ES Dra & \checkmark & $\times$ & \checkmark \\
    LkHA 170 & \checkmark & \checkmark& $\times$\\
    \hline
    \end{tabular}
\end{table*}

\bsp	% typesetting comment
\label{lastpage}
\end{document}